\documentclass[twocolumn,trackchanges]{aastex62}

\usepackage{nccmath}
\usepackage{multirow}
\usepackage{amsmath, amstext}
\usepackage{amssymb}
\usepackage{amsmath}
\usepackage{mathtools}
\usepackage{color} 
\usepackage{threeparttable}
\usepackage{float}

\received{December, 2019}
\revised{January, 2020}
\accepted{\today}

\submitjournal{ApJ}

\shorttitle{Variability of the Mg~II Emission Line and UV Fe~II band in the Blazar CTA~102}
\shortauthors{Chavushyan et al.}

\begin{document}

\title{Flare-like Variability of the Mg~II $\lambda$2798 \AA\ Emission Line and UV Fe~II band in the Blazar CTA~102}

\correspondingauthor{Vahram Chavushyan}
\email{vahram@inaoep.mx, chavushyanv@gmail.com}

\author[0000-0002-2558-0967]{Vahram Chavushyan}
\affil{Instituto Nacional de Astrof\'isica, \'Optica y Electr\'onica, Luis Enrique Erro $\# 1$,
Tonantzintla, Puebla 72840, M\'exico}

\author[0000-0002-5442-818X]{Victor M. Pati\~no-\'Alvarez}
\affil{Instituto Nacional de Astrof\'isica, \'Optica y Electr\'onica, Luis Enrique Erro $\# 1$, 
Tonantzintla, Puebla 72840, M\'exico}
\affiliation{Max-Planck-Institut f\"ur Radioastronomie, Auf dem H\"ugel 69,
D-53121 Bonn, Germany}

\author[0000-0002-9443-7523]{Ra\'ul A. Amaya-Almaz\'an }
\affil{Instituto Nacional de Astrof\'isica, \'Optica y Electr\'onica, Luis Enrique Erro $\# 1$, 
Tonantzintla, Puebla 72840, M\'exico}

\author{Luis Carrasco }
\affil{Instituto Nacional de Astrof\'isica, \'Optica y Electr\'onica, Luis Enrique Erro $\# 1$, 
Tonantzintla, Puebla 72840, M\'exico}

\begin{abstract}

We report on the detection of a statistically significant flare-like event in the Mg~II~$\lambda$ 2798~\AA\ emission line and the UV~Fe~II band of CTA~102 during the outburst of autumn 2017. The ratio between the maximum and minimum of $\lambda$3000~\AA\ continuum flux for the observation period ($2010-2017$) is 179$\pm$15. Respectively, the max/min ratios 8.1$\pm$10.5 and 34.0$\pm$45.5 confirmed the variability of the Mg~II emission line and of the Fe~II band. The highest levels of emission lines fluxes recorded coincide with a superluminal jet component traversing through a stationary component located at $\sim$0.1 mas from the 43 GHz core. Additionally, comparing the Mg~II line profile in the minimum of activity against the one in the maximum, we found that the latter is broader and blue-shifted. As a result of these findings, we can conclude that the non-thermal continuum emission produced by material in the jet moving at relativistic speeds is related to the broad emission line fluctuations. In consequence, these fluctuations are also linked to the presence of broad-line region (BLR) clouds located at $\sim$25 pc from the central engine, outside from the inner parsec, where the canonical BLR is located. Our results suggest that during strong activity in CTA~102, the source of non-thermal emission and broad-line clouds outside the inner parsec introduces uncertainties in the estimates of black hole (BH) mass. Therefore, it is important to estimate the BH mass, using single-epoch or reverberation mapping techniques, only with spectra where the continuum luminosity is dominated by the accretion disk.

\end{abstract}


\section{Introduction} \label{sec:intro}

Since its discovery in 1959 \citep{Harris1960}, the radio source CTA~102 has been studied fairly extensively in the entire electromagnetic spectrum. CTA~102 is a well-known quasi-stellar object, first identified in the optical band by \citet{Sandage1965}, with a redshift of 1.037 \citep{Schmidt1965} showing flux fluctuations in the 32.5 cm band \citep{Sholomitskii1965}. Although it has been classified as an optically violent variable (OVV) QSO \citep{Angel1980}, the long-term B-band monitoring ($1969-1988$) of \citet{Pica1988} shows that its variations are too slow and gentle to fit that classification. The optical linear polarization is strong, sometimes 11\% \citep{Moore1981}. Thus CTA~102 can be classified as a high polarization quasar (HPQ) and a blazar \citep{Nolan1993}. Currently, CTA~102 is well known as a flat-spectrum radio quasar (FSRQ). Furthermore, it demonstrates structural and flux variability on its pc-scale jet \citep{Larionov2013,Fromm2015,Casadio2015,Casadio2019} and also shows correlated variability among different wavebands \citep{Bachev2017,Larionov2016,Raiteri2017,Kaur2018,Li2018,Prince2018,Shukla2018,DAmmando2019}. The extended giant flaring behavior of CTA~102 in late 2016 and early 2017 has permitted the assembly of an exquisite multiwavelength time-resolved database (see papers cited above). Despite the precious database compiled, no consensus about the location of the gamma-ray production zone in CTA~102 has been reached so far. Some studies \citep{Zacharias2017,Kaur2018,Shukla2018} favor the scenario where the gamma-ray emission in CTA~102 is generated close to the central black hole (BH) within or close to the border of the canonical broad-line region (BLR), which is located within the inner pc. However, other works \citep{Zacharias2017,Zacharias2019,Gasparyan2018,Costamante2018} find more feasible the scenario where gamma-rays are produced far from the central BH and canonical BLR, within or downstream of the radio core at distances much larger than 1 pc. In case the latter scenario is correct, a possible source of seed photons is the jet-excited BLR.

The existence of the jet-excited BLR outflowing downstream the jet was proposed to explain the link between jet kinematics on sub-pc scales, optical continuum and emission line variability \citep{Arshakian2010,Leon-Tavares2010}. Direct observational evidence of the BLR close to the radio core of the jet comes from a response of the broad emission lines to changes in the non-thermal continuum emission of the jet \citep{Leon-Tavares2013}. The highest levels of the Mg~II $\lambda$2798 \AA\ emission line and the UV Fe~II band fluxes and gamma-ray outburst happen when a jet component passes through (or is ejected from) the radio core of 3C~454.3. This remarkable event was also confirmed in consequent studies \citep{Isler2013,Jorstad2013}. The differences in slopes of the Baldwin Effect in radio-quiet (RQ) active galactic nuclei (AGN) and FSRQ can be explained by the presence of a second BLR that is related to the jet, probably in the form of an outflow \citep{Patino-Alvarez2016}. The origin of the jet-excited outflow is most likely to be the wind from the accretion disk accelerated to sub-relativistic speeds by a twisted magnetic field of the jet. This possibility was previously suggested by \cite{Perez1989}, and also for specific sources like 3C 273 \citep{Paltani2003} and 3C~454.3 \citep{Finke2010}.  

Relativistic flows may also interact with a cloud from the atmosphere of a red giant star \citep{Bosch-Ramon2012}, or come from a star-forming region that passes through the jet. \cite{Barkov2012b} propose the former scenario to explain the TeV flaring activity in M87. The latter was invoked to explain the short timescale GeV flare-like event generated close to the radio core in 3C~454.3 \citep{Khangulyan2013} and CTA~102 \citep{Zacharias2019}. Previous studies have addressed the dynamics of the interaction of a red giant star after entering an AGN jet \citep[e.g.][]{Araudo2010,Barkov2010,Barkov2012a}; while other works have studied the global impact on the jet propagation and content caused by stars or clouds as they interact with the jet \citep[e.g.][]{Komissarov1994,Steffen1997,Hubbard2006,Choi2007,Sutherland2007,Jeyakumar2009}.

In this work, we explore the variability of the broad emission lines in CTA~102 to use it as an auxiliary piece of information to probe the geometry and physics of the innermost regions of CTA~102 and to provide evidence for the above scenarios of gamma-ray production. Although CTA~102 has been monitored extensively at all wavelengths, the variability of the broad emission lines was not systematically studied. However, some studies \citep{Larionov2016,Bachev2017,Raiteri2017} demonstrated the evolution of the shape of optical spectra and the Mg~II broad-line emission during the flaring events.

The present work is the first one to address the variability of emission lines in CTA~102 during the Fermi/LAT era with the largest sample of its optical spectra ever compiled.

The cosmological parameters adopted throughout this paper are H$_0=71$ km s$^{-1}$ Mpc$^{-1}$, $\Omega_{\Lambda}=0.73$, $\Omega_R=8.3316\times10^{-5}$, $\Omega_m=0.27-\Omega_R$. At the redshift of the source, z=1.037, the spatial scale of 1\arcsec corresponds to a physical scale of 21.2 kpc and the luminosity distance is 6.933 Gpc.

\section{Observations}
\label{sec:observe}

\begin{figure*}[htbp]
\begin{center}
\includegraphics[width=0.9\textwidth]{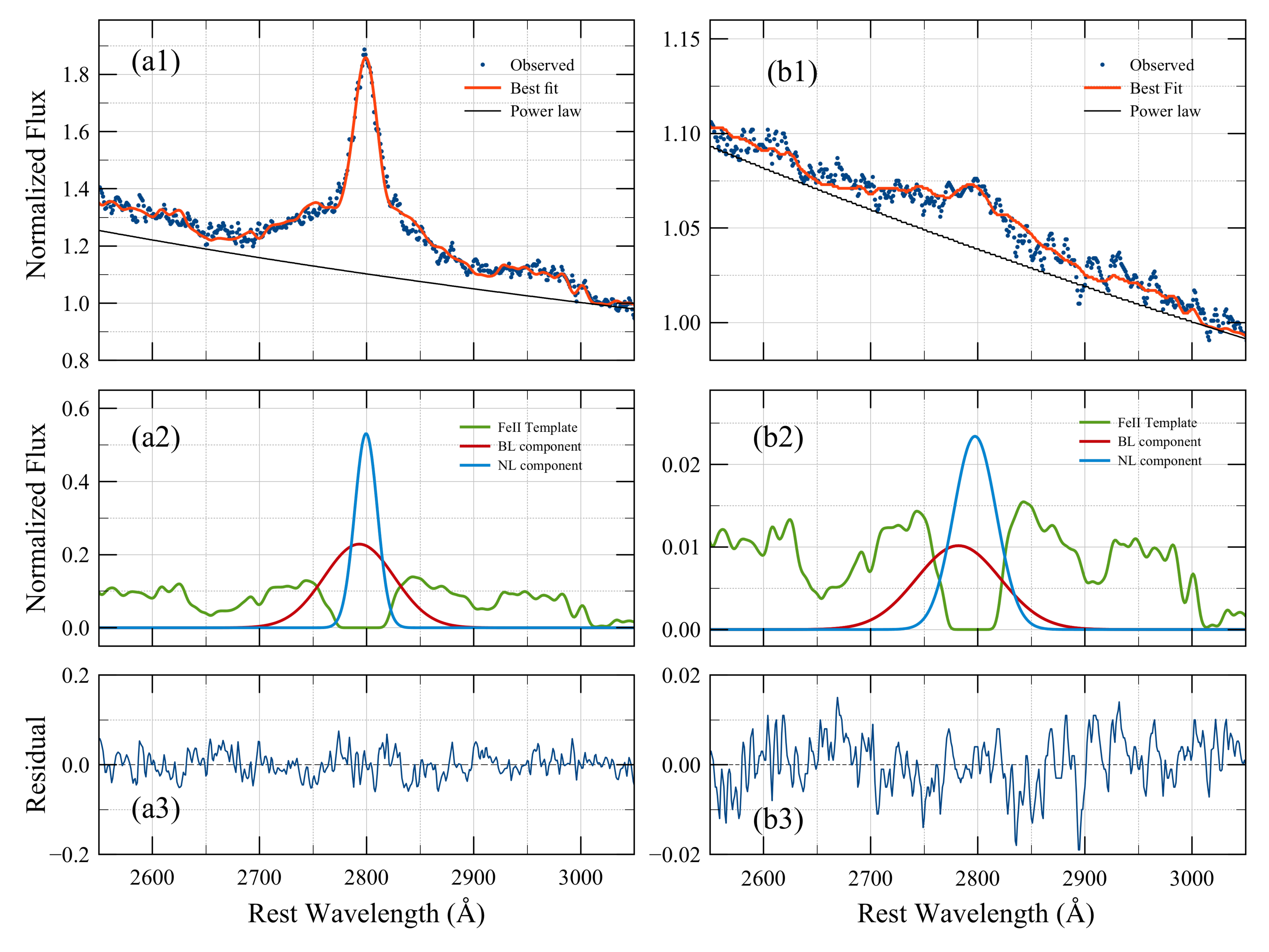}
\caption{Left panel: Decomposition of the continuum, Fe~II, and Mg~II emission line for the spectrum at the minimum continuum flux. Right panel: Decomposition of the continuum, Fe~II, and Mg~II emission line for the best spectrum near the flare of late 2016 - early 2017. (a1, b1) Best fit of the observed spectra and continuum fitting. The spectra are normalized to the continuum flux at $\lambda$3000 \AA. (a2, b2) Fitted spectral components. (a3, b3) Residuals.}
\label{Decomposition}
\end{center}
\end{figure*}

For the analyses performed in this work, we used 271 optical spectra of CTA102 observed in different epochs, calibrated against the V-band magnitude taken from the Ground-based Observational Support of the Fermi Gamma-ray Space Telescope at the University of Arizona monitoring program\footnote{\url{http://james.as.arizona.edu/~psmith/Fermi/}}. The details regarding the observational setup and the process of data reduction can be examined in \cite{Smith2009}. We took the spectra to the rest frame of the object and perform a cosmological correction to the flux of the form $(1+z)^3$. We did not apply correction for galactic reddening.
To accurately measure the flux of the Mg~II $\lambda$2798 \AA\ emission line, an appropriate subtraction of the UV Fe~II band emission, as well as the continuum, is necessary. Here, we present a brief description of our spectral fitting, which is based on using the  \texttt{SPECFIT} task from the \texttt{IRAF} package\footnote{\url{http://iraf.noao.edu/}} \citep{Tody1986,Tody1993}. To subtract the featureless continuum from the spectrum, we approximate it with a power-law function. Next, we removed the contribution of the Fe~II emission by fitting it using the template of \cite{Vestergaard2001}. After performing this procedure to all spectra, we integrated the line profile of the Mg~II emission line from $2740-2860$ \AA\ to measure its flux. The Fe~II emission was measured by integrating the fitted Fe~II template in the wavelength range of $2850-3000$ \AA, where the red UV Fe~II bump is found. The UV continuum flux at $3000$ \AA\ was measured from the iron-subtracted spectrum, by taking the average value between the wavelength range $2950-3050$ \AA; this is done to avoid measurement errors due to noise in the spectrum. Examples of this decomposition process for the spectrum at the minimum continuum flux, and for a spectrum during the flare of late 2016 $-$ early 2017 are shown in Figure~\ref{Decomposition}. Unfortunately, we could not perform an analysis of the Mg~II emission line FWHM evolution because the spectra were taken with different slit sizes, 3\arcsec~being the smallest and 7.6\arcsec~the largest. Hence, this gives a difference in the resolution of $R_{max}\approx2.53 \times R_{min}$.

\begin{figure*}[htbp]
\begin{center}
\includegraphics[width=0.85\textwidth]{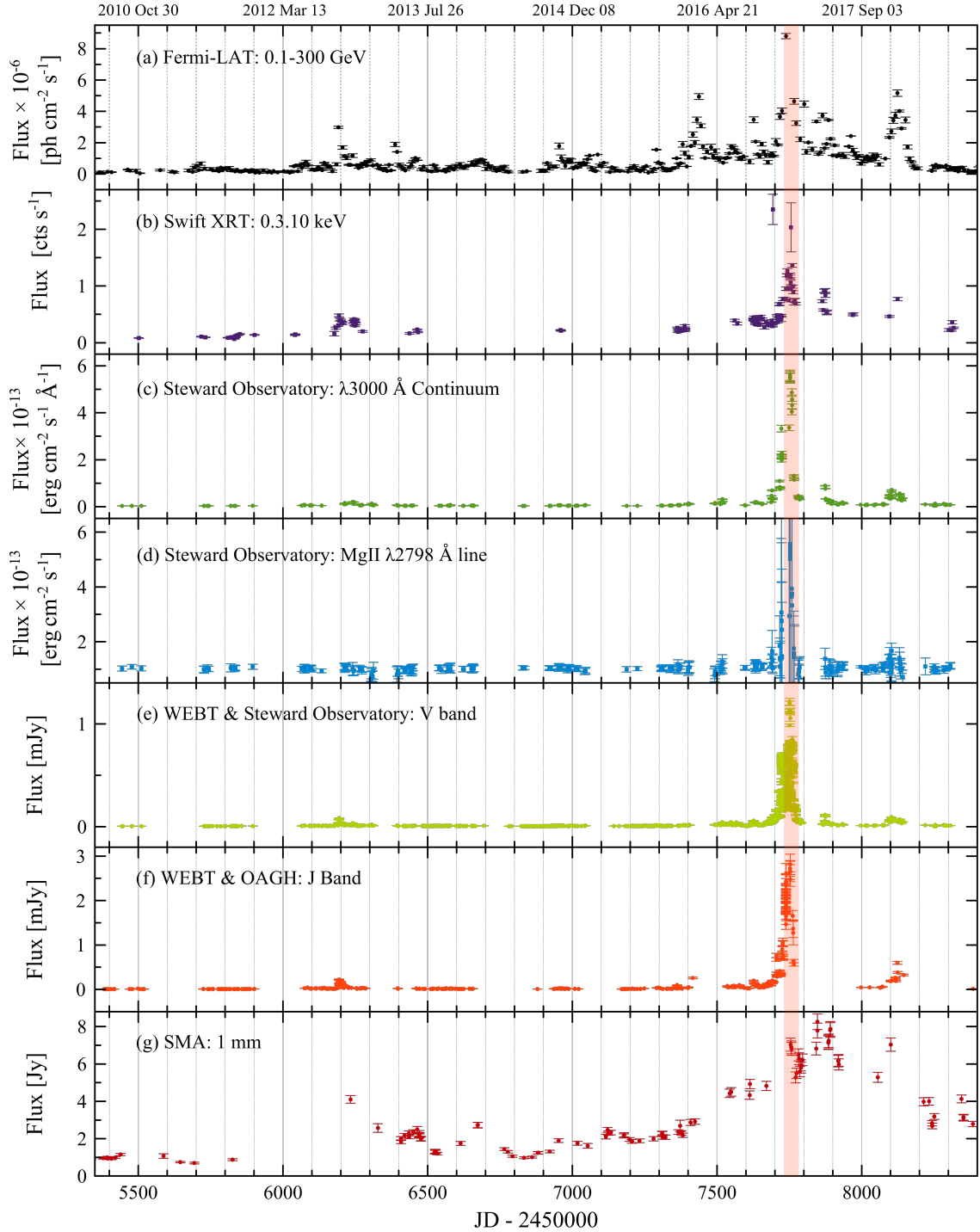}
\caption{Multiwavelength light curves for CTA~102. (a) Gamma-rays light curve derived from Fermi-LAT data in the energy range 0.1-300 GeV. (b) X-rays light curve obtained with Swift-XRT data in the energy range 0.3-10 keV. (c) $\lambda$3000 \AA\ spectral continuum light curve, measured from spectra obtained at the Steward Observatory. (d) Mg~II $\lambda$2798 \AA\ emission line light curve, measured from spectra obtained at the Steward Observatory. (e) Optical V-band light curve built with data obtained at the Steward Observatory and from the WEBT data archive. (f) NIR J-band light curve built with data obtained at the OAGH and from the WEBT data archive. g) 1mm light curve built with data from the public database of the SMA. The red vertical stripes show the time (JD$=2457756.5\pm29$) when the new superluminal blob K1 \citep{Casadio2019} ejected from the radio core passes through the stationary feature C1 \citep{Jorstad2005,Jorstad2017,Casadio2019} in the VLBI (Very Long Baseline Interferometry) maps. Their widths represent the associated uncertainties (taken from the uncertainty in the ejection time from the core).}
\label{multiwavelength}
\end{center}
\end{figure*}

We calculated the error associated with the Mg~II flux measurement accounting for two different contributions. The main contribution is due to the random error produced by the dispersion of the spectra and the signal-to-noise ratio (S/N), estimated following \cite{Tresse1999}.
The second contribution to the error is the one produced by the subtraction of the Fe~II emission, estimated as in \cite{Leon-Tavares2013}. We consider that in the range $2786-2805$ \AA\ no iron subtraction was performed. For the error associated to the Fe~II emission, we are only taking into account the random error, which can be estimated in the same way as that of the Mg~II line \citep[applying the methodology in][]{Tresse1999}. On the other hand, the error in the $\lambda$3000 \AA\ continuum flux is estimated as the rms of the iron-subtracted spectrum around $\lambda$3000$\pm$50 \AA. The measured fluxes and errors  are shown in Table~\ref{table1}. The light curves for continuum emission and Mg~II line are shown in Figure~\ref{multiwavelength} panels (c) and (d).
 
  \begin{table*}[htbp]
\centering
\caption{\tablenotemark{*}Flux measurements for the Mg~II $\lambda$2798 \AA\ emission line, the Fe II UV band, and the $\lambda$ 3000 \AA\ continuum.}
\begin{tabular}{ccccccc}
\hline
\multirow{2}{*}{JD-2450000} & {Mg~II $\lambda$2798 \AA\ Flux} & {Error} & {Fe II Flux} & {Error} & {Continuum $\lambda$3000 \AA\ Flux} & {Error}  \\
 & \multicolumn{2}{c}{$\times10^{-13}\, erg\, s^{-1}cm^{-2}$} & \multicolumn{2}{c}{$\times10^{-13}\, erg\, s^{-1}cm^{-2}$} & \multicolumn{2}{c}{$\times10^{-15}\, erg\, s^{-1}\, cm^{-2}\, {\rm \AA}^{-1}$} \\
\hline
5443.93 &    1.02 &    0.09 &    0.40 &    0.13 &    3.18 &    0.28 \\
5477.76 &    1.09 &    0.09 &    0.43 &    0.12 &    3.47 &    0.26 \\
5510.75 &    1.03 &    0.09 &    0.42 &    0.13 &    3.33 &    0.28 \\
5726.92 &    0.99 &    0.12 &    0.34 &    0.17 &    3.46 &    0.37 \\
5727.91 &    0.98 &    0.13 &    0.33 &    0.18 &    3.44 &    0.39 \\
5738.88 &    1.04 &    0.12 &    0.33 &    0.16 &    3.40 &    0.35 \\
5743.90 &    1.01 &    0.10 &    0.33 &    0.14 &    3.43 &    0.30 \\
5819.91 &    1.04 &    0.15 &    0.41 &    0.20 &    3.57 &    0.45 \\
5831.77 &    1.01 &    0.11 &    0.39 &    0.15 &    3.41 &    0.32 \\
5833.75 &    1.06 &    0.09 &    0.38 &    0.13 &    3.43 &    0.27 \\
\hline
\end{tabular}
\tablenotetext{*}{This table is available in its entirety in a machine-readable form in the online journal.}
\label{table1}
\end{table*}

The gamma-ray data were obtained from the public database of the Large Area Telescope (LAT) on board of the Fermi Gamma-Ray Space Telescope \citep{ Abdo2009}. The weekly light curve in the energy range from 0.1-300 GeV was built by reducing the Fermi-LAT data with the \texttt{Fermitools version 1.0.2}. In the model, we included all sources within 15$^{\circ}$ of the location of CTA 102, extracted from the 4FGL catalog \citep{4FGL}. The light curve is shown in Figure~\ref{multiwavelength} panel (a). The X-ray data were obtained from the public database of the Swift-XRT\footnote{\url{http://www.swift.psu.edu/monitoring/}}. The Swift-XRT data were processed using the most recent versions of the standard SWIFT tools: Swift Software version 3.9, \texttt{FTOOLS version 6.12} \citep{Blackburn1995} and \texttt{XSPEC version 12.7.1} \citep{Arnaud1996}. Light curves are generated using \texttt{xrtgrblc version 1.6} and are shown in  Figure~\ref{multiwavelength} panel (b). Full details of the reduction procedure can be found in \cite{Stroh2013}. The optical V band data were obtained from two different sources, shown in Figure~\ref{multiwavelength} panel (e), the Ground-based Observational Support of the Fermi Gamma-ray Space Telescope at the University of Arizona \citep[Steward Observatory,][]{Smith2009}, and the Whole Earth Blazar Telescope \citep[WEBT,][]{Villata2006}. In  Figure~\ref{multiwavelength} panel (f) are presented the Near-Infrared (NIR) J-band data which were obtained from two different sources, the Observatorio Astrof\'isico Guillermo Haro (OAGH) using the Cananea Near-Infrared Camera \citep[CANICA,][]{Carrasco2017}, and the WEBT project \citep{Villata2006}. The OAGH J-band data was observed using the dithering technique in order to accurately subtract the background, and the photometry calibration was done against magnitudes of bright standard stars found in the Two Micron All-Sky Survey (2MASS). The 1~mm data were retrieved from the Sub-Millimeter Array (SMA) public database\footnote{\url{ http://sma1.sma.hawaii.edu/callist/callist.html}}; shown in Figure~\ref{multiwavelength} panel (g), and full details on the observations and the data reduction process can be found in \cite{Gurwell2007}. 
 
\section{Variability} \label{sec:var}

\subsection{Multiwavelength Variability}

We performed cross-correlation analysis between all the bands studied in this paper. The results can be found in Table~\ref{table2}.

 \begin{table}
\centering
\caption{Cross-correlation results. All delays have correlations at significance level $\geq$ 99\%. All cross-correlations are performed in the order stated in this table.}
\begin{tabular}{ccc}
\hline
Bands & First delay & Second delay  \\
\hline
Gamma-rays vs X-rays & 10$\pm$23 & 322$\pm$23 $^*$ \\
Gamma-rays vs 3000 \AA   & 3.1$\pm$7.1 & -360.0$\pm$7.1 $^*$ \\
Gamma vs Mg~II & 6.9$\pm$7.1 & -360.0$\pm$7.1 $^*$\\
Gamma vs Fe~II & 10.0$\pm$9.7 & -361.7$\pm$9.7 $^*$\\
V-band vs Gamma-rays & -4.1$\pm$7.0 & -314.9$\pm$7.0 $^*$\\
J-band vs Gamma-rays & -1.4$\pm$7.0 & -314.5$\pm$7.0 $^*$\\
Gamma-rays vs 1mm & -42$_{-51}^{+208}$ & --- \\
3000 \AA\ vs X-rays & 18.7$\pm$22.6 & --- \\
Mg~II vs X-rays & 5.3$\pm$22.6 & --- \\
Fe~II vs X-rays & 2.4$\pm$22.6 & --- \\
V-band vs X-rays & 66.4$_{-81.3}^{+40.6}$ & --- \\
J-band vs X-rays & 7.7$\pm$22.6 & --- \\
1mm vs X-rays & 19$\pm$23 & --- \\
Mg~II vs 3000 \AA\ & 0.0$\pm$7.1 & --- \\
Fe~II vs 3000 \AA\ & 0.0$\pm$7.1 & --- \\
V-band vs 3000 \AA & 4.5$\pm$7.1 & -222.7$\pm$52.7 $^{**}$ \\
J-band vs 3000 \AA & -0.7$\pm$7.1 & -210.7$\pm$19.4 $^{**}$ \\
3000 \AA vs 1mm & 2.1$_{-50.9}^{+226.5}$ & --- \\
Fe~II vs Mg~II & 0.0$\pm$7.1 & --- \\
V-band vs Mg~II & 2.4$\pm$7.1 & --- \\
J-band vs Mg~II & 0.0$\pm$7.1 & --- \\
Mg~II vs 1mm & 141.9$_{-122.4}^{+87.4}$ & --- \\
V-band vs Fe~II & 9.5$\pm$9.6 & --- \\
J-band vs Fe~II & 4.9$\pm$9.6 & --- \\
Fe~II vs 1mm &  2.1$_{-52.4}^{+157.3}$ & --- \\
V-band vs J-band & -0.1$\pm$3.1 & --- \\
V-band vs 1mm & -25$_{-231}^{+135}$ & --- \\
J-band vs 1mm & 19.0$_{-52.4}^{+157.3}$ & --- \\

\hline
\end{tabular}
\tablenotetext{*}{~This delay was obtained due to the triple-peaked morphology of the gamma-ray light curve.}
\tablenotetext{**}{~~This delay was found to be an alias.}
\label{table2}
\end{table}

 It is worth noting that most of the cross-correlations involving the 1mm band, result in a large uncertainty in the delay, due to the great difference of the variability timescale and amplitude between the 1mm light curve and the other bands. This can be attributed to the difference in sizes of the emission regions \citep{Leon-Tavares2011}.
 
 We performed an alias check via Fourier analysis, to determine the truthfulness of the delays obtained with the cross-correlation analysis \citep{Press2007}\footnote{\url{https://www.cambridge.org/numericalrecipes}}. We found that the delays obtained around $\sim$200 days in the cross-correlation functions between the $\lambda$3000 \AA\ continuum light curve and the J and V-bands are aliases. On the other hand, we attribute the delays obtained around $\sim$300 days that involve the gamma-ray light curve, to the multiple peaks it shows. These differ from the highest one (in the gamma-rays and in the other bands as well) by $\sim$300 days.

\subsection{Spectral Variability}
\begin{figure}[htbp]
\begin{center}
\includegraphics[width=0.48\textwidth]{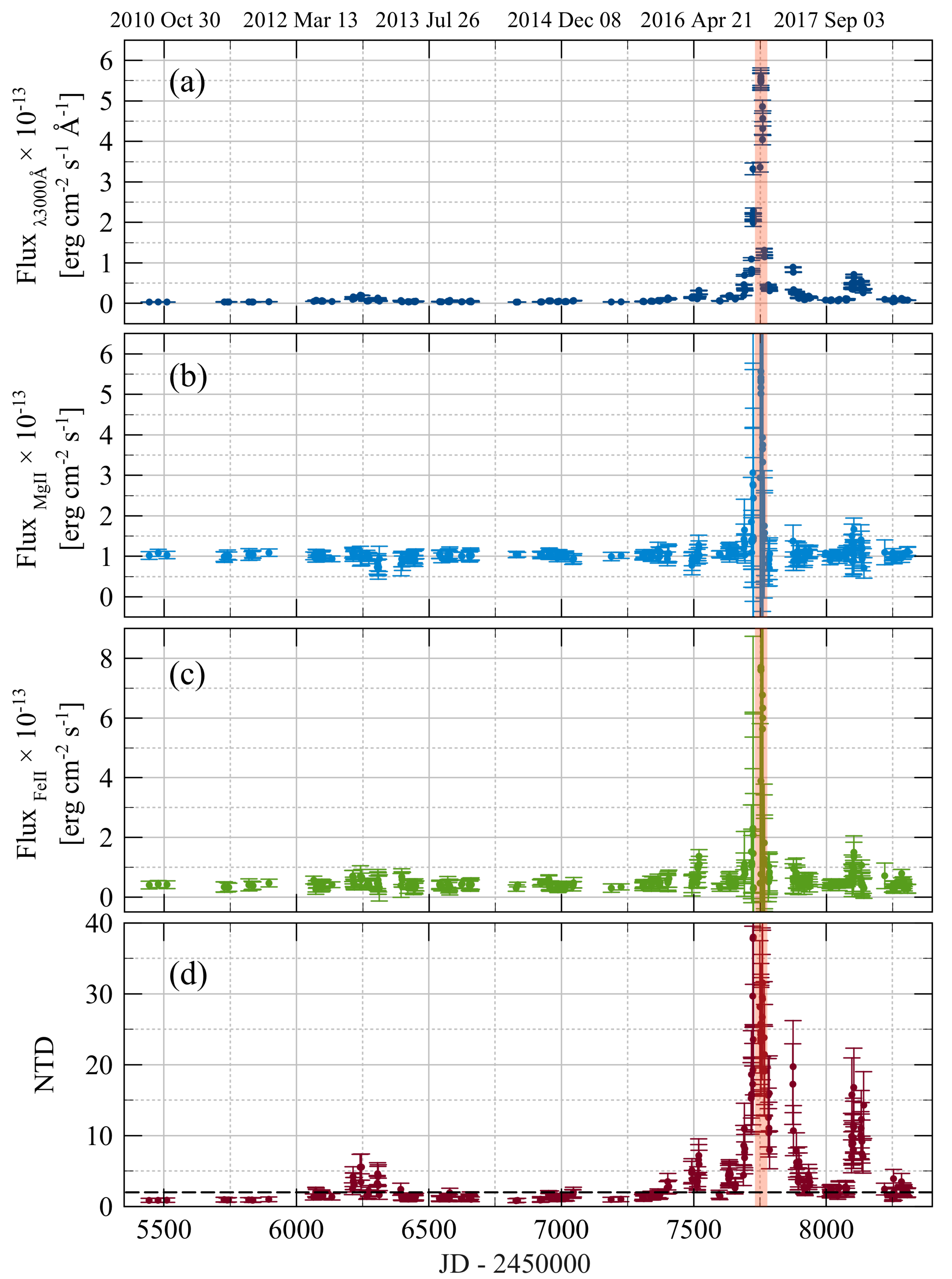}
\caption{Light curves of: (a) the $\lambda$3000 \AA\ continuum flux; (b) the Mg~II $\lambda$2798 \AA\ emission line flux; (c) the UV Fe~II band flux, and (d) the NTD parameter (see definition in Section~\ref{LumRel}). The red vertical lines are the same as in Figure~\ref{multiwavelength}.}
\label{line-continuum}
\end{center}
\end{figure}

The flux light curves for the Mg~II $\lambda$2798 \AA\ line, the UV Fe~II band, and the $\lambda$3000 \AA\ continuum are shown in Figure~\ref{line-continuum} panels (a) through (c). The main focus of this paper is the study of the significant flaring period in late 2016 and early 2017. This flaring period also marks the highest gamma-ray fluxes the source has shown since the Fermi/LAT was launched. The ratio between the maximum and the minimum of continuum flux for the entire observation period is 179$\pm$15, which represents the highest activity for CTA 102 in over a decade. The variability of the Mg~II emission line and the Fe~II band is already notorious from a visual inspection of the light curves. It is confirmed by the large ratios between maximum and minimum flux of 8.1$\pm$10.5 and 34.0$\pm$45.5, respectively. The highest fluxes recorded during the flaring event are at 6.1-$\sigma$ and 6.3-$\sigma$ of the weighted mean for the Mg~II and the Fe~II, respectively. We performed cross-correlation analysis \citep[for details on the cross-correlation methodology applied, see][]{Patino-Alvarez2018} between the Mg~II  emission line, Fe~II, and the continuum  emission. After accounting for possible correlated bias, the resultant lag is 0.0$\pm$7.1 days (uncertainty at 90\%) for all permutations between these three light curves. The uncertainty is the same in all cases, due to all observation dates being the same since they are all measured from the same spectra.

\subsection{Luminosity Relations}\label{LumRel}

\begin{figure}[htbp]
\begin{center}
\includegraphics[width=0.48\textwidth]{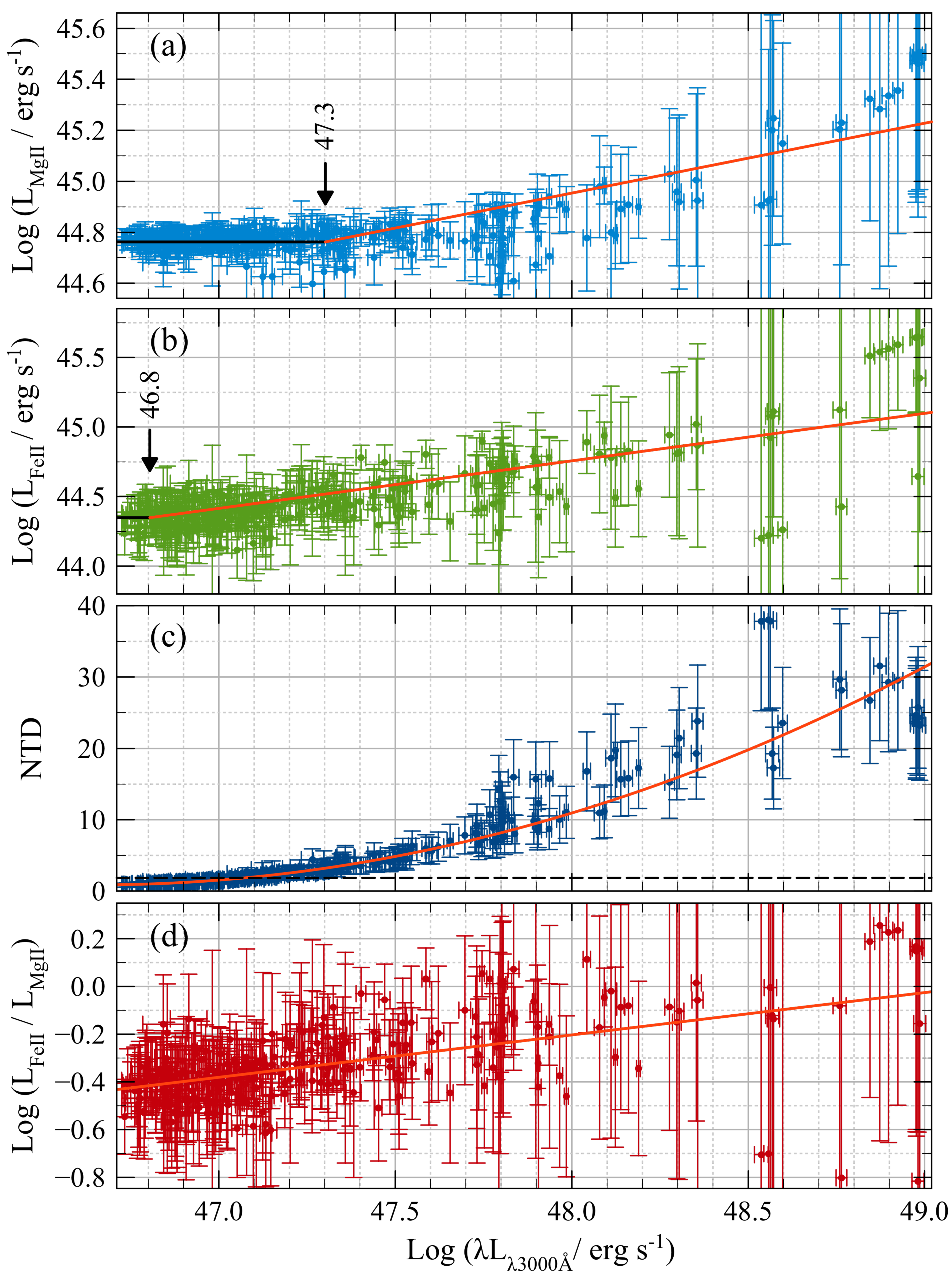}
\caption{(a) Variations of $\lambda$3000 \AA\ continuum luminosity compared to the Mg~II $\lambda$2798 \AA\ emission line 
luminosity. (b) The UV Fe~II band luminosity. (c) The NTD parameter (dashed line represents NTD$=$2). (d) The ratio 
between the UV Fe~II and the Mg~II $\lambda$2798 \AA\ luminosities. The red and black solid lines in panels (a) and (b) 
represent linear fits to the points before and after the thresholds (see Section~\ref{sec:var}), respectively. The solid 
lines in panel (c) and (d) represent a simple fit to the points. All fitted lines are for demonstration purposes only.}
\label{correlations}
\end{center}
\end{figure}

Since the cross-correlation analysis results in a correlation between the continuum emission and both emission features (Mg~II and Fe~II), we look for the relationship between the luminosities, which can be seen in Figure~\ref{correlations}. We tested for correlations between the $\lambda$3000 \AA\ continuum luminosities and the Mg~II $\lambda$2798 \AA\ luminosities, as well as the UV Fe~II band luminosities using the Spearman rank correlation test. Between the continuum and the Mg~II we found a correlation coefficient of 0.48 with a p-value of 2.2$\times$10$^{-14}$, indicating a weak albeit significant correlation; while the test between the continuum and Fe~II yields a correlation coefficient of 0.66 with a p-value of 4.9$\times$10$^{-35}$, indicating a strong and significant correlation. Subsequently, we noted that the luminosities of Mg~II and Fe~II do not increase monotonically as the continuum luminosity increases. There is a luminosity range in which the continuum emission increases, but the Mg~II and the Fe~II seem to fluctuate around a constant luminosity value. Similar behavior can be seen in the blazar 3C 454.3 \citep[see Figure 4,][]{Leon-Tavares2013}.
In order to quantify the luminosity ranges in which the emission features do not vary significantly, we applied an error-weighted linear fit to the logarithmic luminosities using a Levenberg-Marquardt algorithm. First, we fit the 11 points with lower continuum luminosity \citep[11 points is the minimum number required so that a calculated correlation coefficient is statistically meaningful,][]{Alexander1997}. In the case that the slope yielded a result within 1-$\sigma$ from zero (indicating no correlation), we add the next point sorted by continuum luminosity, until the resulting slope is outside 1-$\sigma$  of zero. This marks the threshold when the Mg~II and Fe~II luminosities started correlating with the continuum luminosity.

These correlations were further tested by applying the Spearman correlation rank test, to the points before and after this threshold, separately. For the Mg~II emission line the threshold lies at $\log{(\lambda L_{\lambda 3000} / erg \, s^{-1})}=47.3$. The Spearman rank correlation yields a correlation coefficient of 0.06 along with a p-value of 0.48 for the points before this threshold, indicating no correlation. While it yields a correlation coefficient of 0.67 with a p-value of $5.7\times 10^{-15}$ for the points after the threshold, indicating a strong and significant correlation (we consider a correlation as significant when the chance of obtaining the correlation coefficient by chance is 5\% or less, i.e. p-value $\leq$ 0.05). For the UV Fe~II band the threshold is found at $\log{(\lambda L_{\lambda 3000} / erg \, s^{-1})}=46.8$. The Spearman rank correlation yields a correlation coefficient of 0.36 and a p-value of 0.14 for the points before the threshold, indicating a weak and not significant correlation. While for the points after the threshold, the correlation is strong and significant with a correlation coefficient of 0.66 and a p-value of $1.0\times 10^{-32}$.

Now that we have shown that the variability of Mg~II and Fe~II in CTA~102 is statistically significant, it is instructive to investigate if there is a causal link between major emission lines flux variations and variability at other wavelengths and their nature.

In order to determine whether the continuum flare has a thermal or non-thermal origin, we calculated the Non-Thermal Dominance parameter, NTD \citep[see][]{Shaw2012}. An expanded definition of NTD for FSRQ is presented by \cite{Patino-Alvarez2016}. The authors define alternative NTD for FSRQ as $NTD=L_{obs}/L_{pred}=(L_{disk}+L_{jet})/L_{pred}$, where $L_{obs}$ is the observed continuum luminosity, $L_{pred}$ is the predicted disk continuum luminosity estimated from the emission line luminosity, $L_{disk}$ is the continuum luminosity emitted by the accretion disk, and $L_{jet}$ is the jet contribution to the continuum luminosity. If the emission line is only ionized by the disk $L_{pred}= L_{disk}$, so that $NTD=1+L_{jet}/L_{disk}$, which shows that $NTD\geq1$. If the $NTD = 1$ it means that the continuum is due only to thermal emission, $1 < NTD < 2$ shows that a superluminal jet exists that contributes to the continuum luminosity, and $NTD > 2$ means that $L_{jet} > L_{disk}$. We estimated $L_{pred}$ by applying the bisector fit to the relationship between the luminosity of the Mg II, and continuum luminosity found by \cite{Shen2011} for a non-blazar sample (see Figure~\ref{Lc-Lmg}). The NTD light curve is shown in Figure~\ref{line-continuum} panel (d).  

During the flare the NTD reached values up to 38, which is a clear indication that the continuum flare was dominated by emission from the jet (non-thermal synchrotron emission). 

The 1 mm light curve also shows an increase in activity during the flaring period in the emission line and continuum, as well as the gamma-rays and X-rays (Figure~\ref{multiwavelength}). This, along with the strong variability of the polarization fraction in the jet \citep{Raiteri2017}, points to the flaring event being caused by non-thermal processes. Moreover, \cite{Casadio2019} found that during the multiwavelength outbursts, a component that was ejected from the core in July 2016 was crossing another stationary feature located very close to the core ($\sim$0.1 mas) increasing its flux density; which further supports the hypothesis that the multiwavelength outburst was caused by non-thermal processes in the jet.

 \subsection{Disk-dominated light curves}
 
Given the predominance of the jet emission during the flaring event, we decided to analyze separately the parts of the spectroscopic light curves where the accretion disk is expected to be the dominant source of continuum emission, i.e., when $NTD<2$. This was done under the assumption that if we analyze these time-ranges, then the results from the cross-correlation would mostly reflect the relationship between the accretion disk and the canonical broad line region. In an attempt to avoid the possible bias from the disk dominated (DD) light curves to exhibit alias, due to the continuum and emission features being measured from the same spectrum, we identified the time ranges where the $NTD<2$ in the spectra. Afterwards, we selected the V-band flux observations that lie between these time ranges, and performed cross-correlation between the DD V-band light curve, and the DD Mg~II and Fe~II light curves. We also performed cross-correlation between the DD V-band and the DD $\lambda$3000 \AA\ continuum, to ensure that their behavior is as close as expected. We obtained a delay of $0.7\pm15.8$ days, well within the range of simultaneous variability. We did not find a significant correlation between the DD V-band light curve and the DD Mg~II line light curve, nor with the DD UV Fe~II band.

\subsection{Line Profile Variability}

All the spectra in the peak of the flare were taken using a slit width of 7.6\arcsec~and a exposure time of 240 seconds, reflecting in the spectra by being of low resolution and having very poor S/N ratio, respectively. We looked for spectra in the minimum flux taken with the same resolution, to compare the profiles at these activity levels, and were lucky to find a spectrum taken with this particular slit width, this being the only one in the entire dataset.

\begin{figure*}[htbp]
\begin{center}
\includegraphics[width=0.9\textwidth]{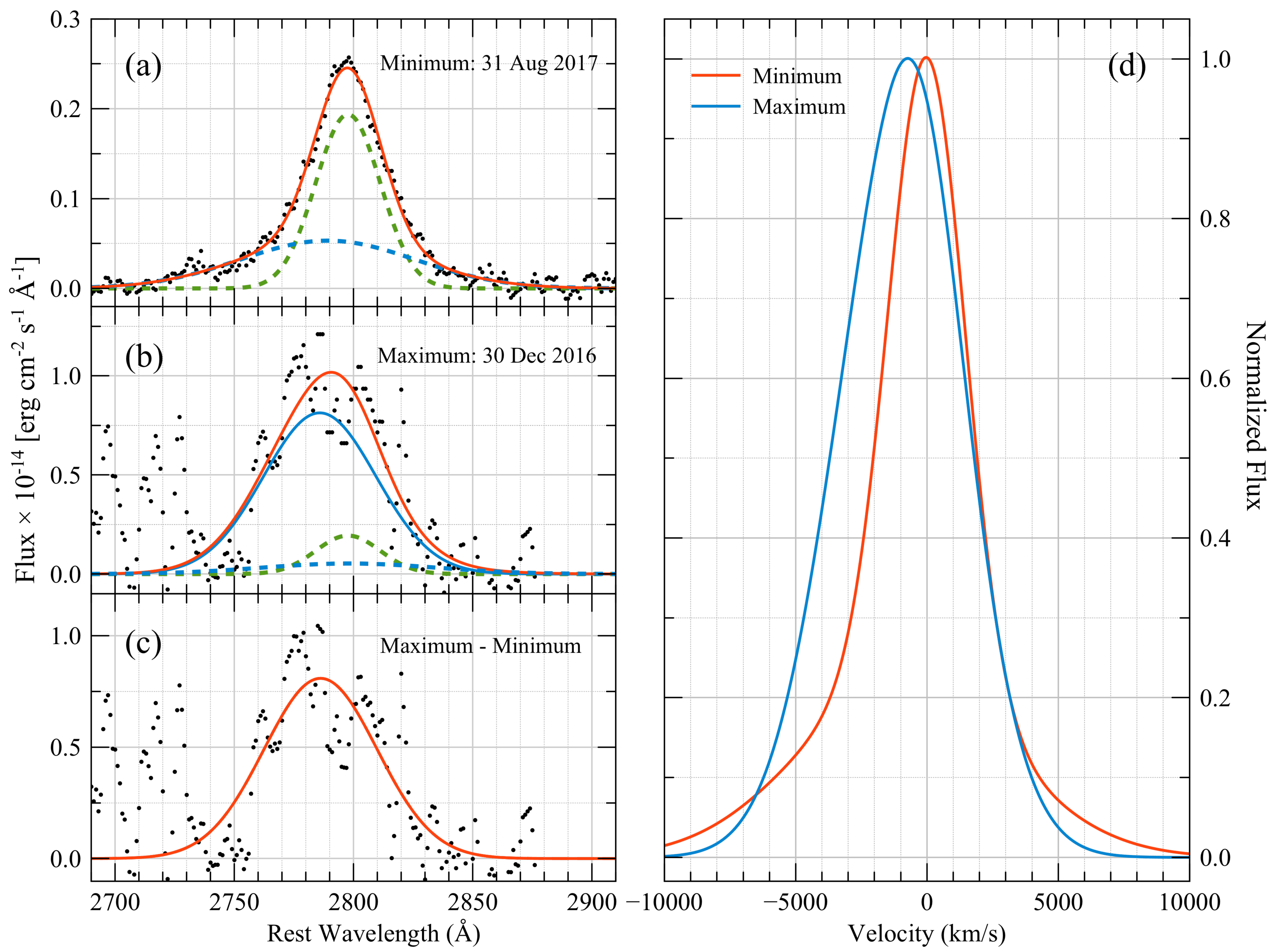}
\caption{(a) The Mg~II $\lambda$2798 \AA\ line profile (black dots) in the minimum flux. The red solid line represents the best fit to the line profile while its two broad and narrow Gaussian components are represented by the blue and green dashed lines, respectively. (b) The Mg~II line profile in the maximum flux, this being the best spectrum in the peak of the flare event. The same two components used to fit the profile in the minimum flux are fixed and a third Gaussian component (blue solid line) were used to fit the line profile in the flare, represented by the red solid line. (c) The resulting profile (black dots) from the subtraction of the profile in the minimum flux from the profile in the maximum flux. The red solid line represents the Gaussian used to fit the resulting profile. (d) The best fit curves to the Mg~II $\lambda$2798 \AA\ line profile in the maximum and minimum of activity normalized to their respective maximum value.}
\label{Line Profile Variability}
\end{center}
\end{figure*}

If we assume that the accretion disk activity is not changing, it means that the two components used to fit the profile in the minimum flux do not change either. Under this assumption, we approximated the additional component we argue is related to jet activity.
The FWHM of the components used to fit the Mg~II line profiles were estimated as upper limits since we need the correction for instrumental profile to determine the value precisely and this information is not publicly available. In Figure~\ref{Line Profile Variability} panel (a), the narrow and broad components (referring to the relative width of the components, instead of the emission line region) from the minimum of activity show FWHM $< 1700 ~km~s^{-1}$ and FWHM $< 4600~km~s^{-1}$, respectively. The additional component needed to fit the profile in the maximum of activity shows a FWHM $<$ 2900 km/s. In Figure~\ref{Line Profile Variability} panel (a), it is evident that the broad component is blue shifted ($\Delta$V $\simeq$ $-900 ~km~s^{-1}$) and, in the panel (b), the third component is also blue shifted ($\Delta$V $\simeq$ $-1200 ~km~s^{-1}$). For the same reason as for FWHM, the shifts can not be determined with precision. The best fit to the resulting profile from the subtraction of the profile in the minimum flux to the profile in the maximum flux, shown in Figure~\ref{Line Profile Variability} panel (c), is assumed to be the third component needed to fit the profile in the maximum of activity. We can speculate that this additional component is the one responsible for the flare in the Mg~II emission line and hence, comes from an additional emission region.

The total Mg~II line profiles in the maximum and minimum were normalized to be able to compare the profile shape during the two different states of activity is presented in Figure~\ref{Line Profile Variability} panel (d). The fit corresponding to the maximum flux seems to be blue shifted ($\Delta$V $\simeq$ $-700 ~km~s^{-1}$) and broader than the one in the minimum flux.

It was not possible to analyze and make conclusions regarding the shape and displacement of the Fe~II profile, in the same way as for the Mg II line profile, since the spectra do not have sufficient resolution and, due to big errors.
 \\

\section{Discussion} \label{sec:disc}

\subsection{Mg~II and Fe~II variability}

The fact that the Mg~II $\lambda$2798 \AA\ and UV Fe~II band flaring events occurred at a time where the continuum was being dominated by non-thermal emission, suggests that the ionization source for the broad-line region material responsible for this emission line event is the jet itself. 

The variability of the Mg~II emission line has been subject of study before. \cite{Clavel1991} concluded that the amplitude of variation of a given emission line seems to depend on the degree of ionization of its ion, by studying for 8 months the variability in NGC~5548 with the International Ultraviolet Explorer (IUE). The highest ionization lines show the strongest variations. On the other hand, the Mg~II, the lowest ionization line displays the smallest amplitude fluctuations. This might explain why the DD Mg~II light curve did not show a significant correlation with the continuum emission. In the SWIFT/UVOT grism monitoring of NGC~5548, \cite{Cackett2015} showed that the Mg~II emission line is not strongly correlated with the continuum variability, and there is no significant lag between the two. However, more recent studies of CTS~C30.10 (radio quiet quasar) and 3C~454.3 (blazar) \cite{Czerny2019} and \cite{Nalewajko2019} show that there are time delays of $562^{+116}_{-68}$ and $\sim$600 days, respectively, between the continuum variations and the response of the Mg~II emission line.

\cite{Leon-Tavares2013} and \cite{Isler2013} reported the statistically significant flare-like event in the Mg~II emission line in the blazar 3C~454.3. This flare coincides with a superluminal jet component traversing through the radio core and correlated with the flares in gamma-rays, UV continuum, and R band flares \citep{Leon-Tavares2013}. It is important to mention that during the flare-like event (brightest flare in 3C~454.3) in the UV continuum in the middle of 2014, \cite{Nalewajko2019} did not identify short-term flaring activity of the Mg~II line flux, unlike the previously reported event.

The physics of broad emission lines with high ionization and high excitation is relatively simple, and they are in good agreement with the observations. However, this does not apply to some lines with low ionization, especially Fe$^+$ lines. These lines are observed in most type 1 AGN in the infrared, visible and ultraviolet spectrum. Currently, there is a disagreement between theoretical calculations assuming photoionization by the central source, with the observations. For more details about the physics of  Fe$^+$  emission lines see \cite{Netzer2013}.

The variability behavior of the optical Fe~II emission in radio-quiet AGN, specifically in Seyfert galaxies, is well studied but not yet well understood \citep[e.g.][and references therein]{Shapovalova2012, Barth2013}. \cite{Shapovalova2012} found that the correlations between the continuum and emission lines are weak, whereas between the optical Fe~II and the continuum is slightly higher and more significant. The authors also conclude that the correlation between the continuum flux and the different Fe~II emission of groups depends on the type of transition. \cite{Barth2013} showed for NGC~4593 and Mrk~1511 highly correlated variations between Fe~II and continuum emission. From reverberation lags, the authors demonstrate that the Fe~II emission in these galaxies originates in photoionized gas, located predominantly in the outer portion of the broad-line region. These results coincide with ours, in that the Fe~II correlates better with the continuum emission than the Mg II emission line does, for the full data set (see Sec.~\ref{LumRel}).

A recent study of the UV Fe~II line complex over the 2000-3000\AA\ region of NGC~5548 is presented by \cite{Cackett2015}. The authors found that the UV Fe~II light curve does not show any clear variability correlated with the continuum light curve. This coincides with our result for the DD light curves.

The flare-like variability of UV Fe~II emission for a blazar type source (3C~454.3) was reported for the first time in \cite{Leon-Tavares2013}, which resulted to be very similar to the Mg~II flare.

\subsection{Implications of the luminosity relations}

Regarding the behavior of the Mg~II emission line and Fe~II, at luminosities lower than the aforementioned luminosity thresholds, \cite{Cackett2015} found that the source NGC 5548 behaves in a similar manner. When the continuum luminosity increases, the emission features fluxes remain relatively constant. On the other hand, \cite{Shapovalova2008} found in the source NGC 4151, that for low continuum fluxes, there is a correlation between the optical continuum emission and optical emission lines (H$\gamma$, He~II, H$\beta$ and H$\alpha$). However, for high continuum fluxes, there is no correlation between the continuum emission and any of the emission lines, with the emission line fluxes remaining relatively constant. One possible interpretation for the emission line behavior in CTA~102, and the aforementioned sources, is to consider that the Mg and Fe contained within the canonical BLR, is already fully ionized or becomes fully ionized (in the case of NGC 4151). This would mean that even when the rate of ionizing photons is increased, the total amount of emission line photons emitted by the canonical BLR does not change significantly. 

We cannot exclude the possibility that the variability is weak and is not detected due to relatively large errors. In this case, the other interpretation is that the clouds of Mg~II and Fe~II are close to the external physical limit of the canonical BLR \citep[e.g.][and references therein]{Barth2013,Guo2019}. \cite{Guo2019} show that the observed weak variability of the Mg~II line comes naturally from photoionization calculations that capture various excitation mechanisms, radiative transfer effects, and variability dilution due to the larger average distances the Mg~II gas reaches. All these effects likely also contribute to the difference between the Mg~II and the Balmer lines variability. For either of the two possibilities, the accretion disk is assumed to be the main source of ionizing continuum, which means that these scenarios are only valid for times when $NTD<2$.

As concluded before, the large flaring event in the emission features is considered to be related to the jet. This tells us that the Mg~II and Fe~II luminosities, after their respective continuum luminosities thresholds, can be assumed to be dominated by jet activity, just as the continuum emission is. This would explain comprehensively why is there a region where there is no significant change in the emission features while the continuum increases. This can be interpreted as the disk-dominance zone (including the canonical BLR); while after the threshold we have the jet-dominance zone. We assess this by looking for a correlation between the NTD parameter and the continuum luminosity. The Spearman rank correlation test yields a correlation coefficient of 0.99 with a probability of obtaining this value by chance, of zero (to machine accuracy), indicating that the continuum luminosity increases, as the jet-dominance increases. Figure~\ref{correlations} panel (c) shows the comparison between the continuum luminosity and the NTD parameter, in which we can see that for low continuum luminosities, the NTD has values lower than 2, suggesting that the accretion disk is the dominant source of the continuum. While at continuum luminosities near the continuum thresholds, we obtain $NTD>2$, suggesting that the found thresholds indeed indicate a change in regime between disk-dominance and jet-dominance.

These results open the question of why the continuum luminosity thresholds between Mg~II and Fe~II are so different, considering that their ionization energies are similar (7.64 eV for the Mg and 7.9 eV for the Fe), which means that the ionizing continuum capable of ionizing Fe is also capable of ionizing Mg. The first possible explanation is that there are simply more Fe atoms than Mg atoms. Therefore, the probability of ionizing a Fe atom is higher, which will cause it to increase its emission with a lower rate of ionizing photons than the Mg would need, due to the lower quantity of available atoms. This could also explain the effect observed in figure~\ref{correlations} panel (d), the Fe~II luminosity increases faster than the Mg~II line luminosity (Spearman rank correlation coefficient of 0.58 with a p-value of 1$.2\times10^{-25}$). Therefore, a higher number of Fe~II photons are emitted as the ionizing continuum increases, thus more Fe atoms are being ionized than Mg atoms. This possibility was studied by \cite{Verner2003}, by using an 830 level model atom for Fe~II in photoionization calculations, and reproducing the expected physical conditions in the BLR of quasars. The modeling reveals that interpretations of high values of UV Fe~II/Mg~II are sensitive not only to Fe and Mg abundance, but also to other factors such as micro-turbulence, density, and properties of the radiation field. 

As mentioned above, we conclude that the flaring behavior in the Mg~II and Fe~II, is produced in a region separated from the canonical BLR. The matter to solve is the location of said emission region. It has been shown that the continuum emission flaring behavior is caused by the interaction between a moving jet component at pc-scales \citep[K1,][]{Casadio2019} with a stationary feature \citep[C1,][]{Jorstad2005,Jorstad2017,Casadio2019}. This was interpreted as a recollimation shock, which can trigger both radio  \citep{Fromm2013a} and gamma-ray outbursts  \citep{Casadio2015}. The cross-correlation analysis shows a delay of 0.0$\pm$7.1 days, between the continuum flux and the emission features, which means that the line emission production zone is nearly co-spatial with the source of the continuum emission. Hence, the additional line emission region, active during the studied flaring event, is close or around the stationary component C1, at $\sim$ 0.1 mas from the 43 GHz core \citep{Jorstad2005,Jorstad2017,Casadio2015}. It translates to a projected distance of 2.12 pc \citep[de-projected distance of 18 pc at a viewing angle of 2.6$^{\circ}$,][]{Fromm2013b}. \cite{Fromm2015} estimated the distance from the 43 GHz core to the black hole, from the core shift measurements, yielding a distance of $7.0 \pm 3.2$ pc. This means that the stationary component C1, which we speculate is associated with an additional BLR component, is located at $\sim$25 pc from the black hole and far from the expected sizes for the canonical BLR.
 \\

\begin{figure}[htbp]
\begin{center}
\includegraphics[width=0.47\textwidth]{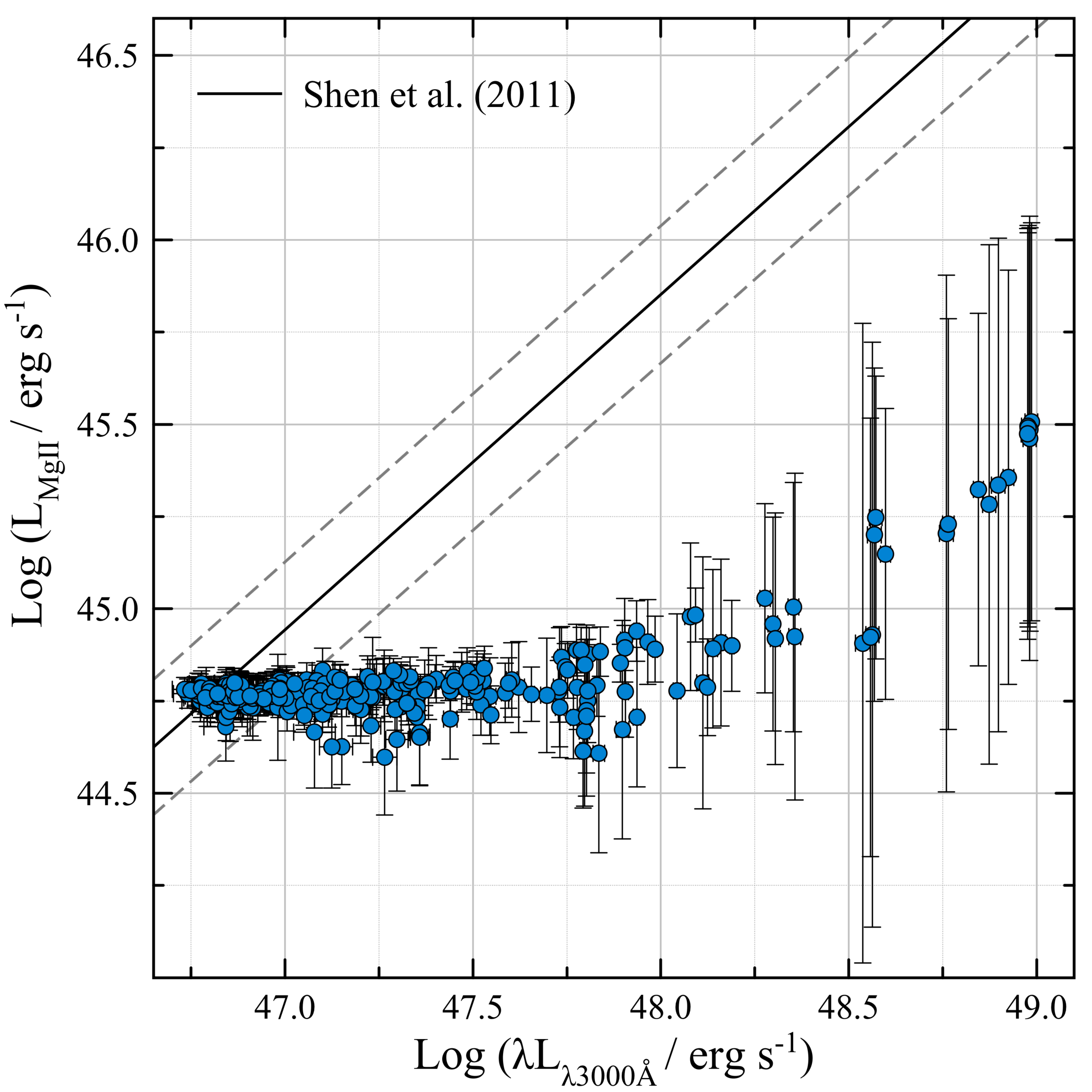}
\caption{The blue dots represent variations of the Mg~II $\lambda$2798\AA\ emission line luminosity compared to the $\lambda$3000\AA\ continuum luminosity, for CTA 102 (our data). The black solid and dashed lines represent the bisector fit and 1-$\sigma$ errors, respectively, for non-blazar sample of \cite{Shen2011}.}
\label{Lc-Lmg}
\end{center}
\end{figure}

In an attempt to assess how the jet presence affects the luminosity relation, we compared our data to a well-defined correlation between the luminosity of the Mg~II emission line to the $\lambda L_{\lambda 3000}$ for the non-blazar sample of \cite{Shen2011}, as it is illustrated in Figure~\ref{Lc-Lmg}. If the optical continuum of strong radio-loud sources is significantly increased by the emission of a jet, its $\lambda L_{\lambda3000}$ (and therefore the virial mass) will be systematically overestimated. If jet contamination is important, we would expect that strong radio objects remain systematically to the right of the best fit relationships for radio-quiet objects (black line) shown in Figure~\ref{Lc-Lmg}. As was shown by  \cite{Greene_Ho2005} for the $\lambda L_{\lambda 5100} - L_{H\beta}$ relation, strong radio sources do not show a significant displacement from the radio-quiet sample. However, our data show a significant displacement concerning the non-blazar sample and are to the right of the relation of  \cite{Shen2011}. When the jet activity is higher (see panels (d) and (c) in Figures~\ref{line-continuum} and \ref{correlations}, respectively), the difference is more noticeable.
 
A remaining unknown that follows is the nature of the additional BLR component at $\sim$25 pc from the black hole. The existence of a dynamic and extended BLR has been suggested before \citep{Popovic2001,Elitzur2006}, while the existence of an outflowing BLR excited by the jet was previously proposed as an explanation for the link between optical continuum and emission lines with observed jet kinematics on sub-pc scale in other sources \citep{Arshakian2010,Leon-Tavares2010,Leon-Tavares2013}. A wind from the accretion disk later accelerated by magnetic fields in the jet, is a likely origin for the jet-excited outflow.

The other scenario widely studied is the one in which the relativistic flows may also interact with a cloud coming from the atmosphere of a red giant star \citep{Bosch-Ramon2012}, or come from a star-forming region that passes through the jet.
 
\section{Summary} \label{sec:sum}

The blazar CTA~102 is part of a sample of objects in the Ground-based Observational Support of the Fermi Gamma-ray Space Telescope at the University of Arizona monitoring program, which allowed us to analyze its behavior over seven years ($2010-2017$). The redshift of this particular object ($z=1.037$) allows us to access the near-UV region of the spectrum. Therefore, allowing us to analyze the light curves of the Mg~II $\lambda$2798 \AA\ emission line, the UV Fe~II band, and the corresponding $\lambda$3000 \AA\ continuum. The results of this work are summarized as follows:

\begin{enumerate}

\item We find a statistically significant flare-like event in the Mg~II emission line and the Fe~II band. The ratio between the maximum and the minimum of continuum flux for the entire observation period is 179$\pm$15. The variability in Mg~II and  Fe~II show $max/min$ ratios of 8.1$\pm$10.5 and 34.0$\pm$45.5, respectively. The highest fluxes recorded during the flaring event are at 6.1-$\sigma$ and 6.3-$\sigma$ of the weighted mean for Mg~II and Fe~II, respectively. The Mg II line profile in the maximum flux seems to be broader and shifted to the blue (with an apparent blue asymmetry), compared to the Mg II profile in the minimum flux.

\item The highest flux levels in the continuum and emission features coincide with the time when a superluminal moving jet component was crossing a recollimation shock (stationary component). The fact that the Mg~II and Fe~II flaring events occurred at a time when the continuum was being dominated by non-thermal emission, suggests that the ionization source for the broad-line region material responsible for this emission line event is the jet itself.

\item The luminosities of the Mg~II line and the Fe~II band do not increase monotonically as the continuum luminosity increases, but rather, there is a luminosity range in which the continuum emission increases while the flux of the emission features oscillates around a constant value. We found the threshold value for continuum emission after which there is a strong and significant correlation between the continuum and the emission features. The threshold lies at $\log{(\lambda L_{\lambda 3000} / erg \, s^{-1})}=47.3$ and $46.8$ for Mg~II and Fe~II respectively. Since we concluded that the flaring event in the Mg~II and Fe~II is jet-related, we can infer that the emission at continuum luminosities higher than the thresholds is dominated by the jet as well.

\end{enumerate}

These results prove that there is a relation between the non-thermal continuum from the jet, and the variability of broad emission lines; suggesting the existence of BLR material at a distance of $\sim$25 pc from the central black hole (BH). This jet-excited BLR may also serve as a source of seed photons for high energy production along the jet by means of external inverse Compton, during flaring activity in emission lines.

The Mg~II emission line study of \cite{Popovic2019} for a representative sample of 284 AGN type 1, found that Mg~II is more complex in nature compared to H$\beta$. It seems to be composed of two components, one virialized, and the other, related to a region of outflows-inflows. The last one contributes to the emission of the Mg~II broad line wings.  The authors suggest that the width at half the maximum of Mg~II can be used to estimate the mass of BH when the virial component is dominant and care should be taken when the Mg~II FWHM $ >  6000~km~s^{-1}$ and/or in the case of strong blue asymmetry. In the most recent study of a sample of 16 extreme variability quasars (EVQs), \cite{Yang2019} showed that the Mg~II emission line varies in the same direction than the continuum flux. However, line width (FWHM) is not varying accordingly with continuum flux variations in most sources \citep[see also,][]{Homan2019}. They conclude, that estimating the BH mass via the width of the broad Mg~II emission line introduces bias due to luminosity dependence. 

The fact that the BLR at a distance of $\sim$25 pc from the BH in CTA~102 responds to changes in the non-thermal continuum makes it difficult to estimate the virial mass of BH from a single-epoch \citep{Vestergaard_Peterson2006} or by applying the reverberation mapping technique \citep{Blandford_McKee1982,Peterson1993}, for the Mg II emission line \citep{Wang2009} in the case of our source. These mass estimation methods depend on there being a single source of ionization (the accretion disk) and virial equilibrium of the BLR clouds. These assumptions cannot be fulfilled during episodes of strong activity in CTA 102, therefore, the ionization source of non-thermal emission and the existence of BLR clouds outside the internal parsec introduces uncertainties in the estimates of BH mass. In this regard, the authors suggest that to obtain the BH mass in blazar-type AGN, using single-epoch or reverberation mapping techniques, only spectra where $NTD<2$ (continuum luminosity dominated by the accretion disk) should be used for the estimations.

\acknowledgments 

Acknowledgments. We thank the anonymous referee for their comments and suggestions that helped to improve the paper. This work was supported by CONACyT (Consejo Nacional de Ciencia y Tecnolog\'ia) research grant 280789 (M\'exico). RAA-A acknowledges support from the CONACyT program for PhD studies. Data from the Steward Observatory spectropolarimetric monitoring project were used. This program is supported by Fermi Guest Investigator grants NNX08AW56G, NNX09AU10G, NNX12AO93G, and NNX15AU81G \url{http://james.as.arizona.edu/~psmith/Fermi/}. The optical and near-infrared photometry partly based on data collected by the WEBT collaboration and stored in the WEBT archive at the Osservatorio Astrofisico di Torino - INAF (http://www.oato.inaf.it/blazars/webt/); for questions regarding their availability, please contact the WEBT President Massimo Villata ({\tt massimo.villata@inaf.it}). 1~mm flux density light-curve data from the Submillimeter Array was provided by Mark A. Gurwell. The Submillimeter Array is a joint project between the Smithsonian Astrophysical Observatory and the Academia Sinica Institute of Astronomy and Astrophysics and is funded by the Smithsonian Institution and the Academia Sinica. The authors thank A. Porras, E. Recillas and G. Escobedo for their help in the observations as part of the Blazar NIR Monitoring Program at OAGH. \\

\software{IRAF \citep{Tody1986,Tody1993},
FTOOLS \citep{Blackburn1995},
XSPEC  \citep{Arnaud1996}, 
Fermitools (v 1.0.20)
}

\bibliographystyle{aasjournal}  

\bibliography{CTA102_refs} 

\end{document}